\documentclass{osa-article}

\journal{osac}
\articletype{Research Article}
\begin{document}

\title{Cross-polarized surface lattice resonances in a rectangular lattice plasmonic metasurface}

\author{M.~Saad~Bin-Alam,\authormark{1,*} Orad~Reshef,\authormark{2} Raja~Naeem~Ahmad,\authormark{2,3} Jeremy~Upham,\authormark{2} Mikko~J.~Huttunen,\authormark{4} Ksenia~Dolgaleva,\authormark{1,2} and Robert~W.~Boyd\authormark{1,2,5}}

\address{\authormark{1}School of Electrical Engineering and Computer Science, University of Ottawa, Ottawa, ON \ K1N 6N5, Canada\\
\authormark{2}Department of Physics, University of Ottawa, Ottawa, ON \ K1N 6N5, Canada\\
\authormark{3}Max Plank Institute of Quantum Optics, Hans Kopfermann Str 1, 85748, Garching bei Munich, Germany\\
\authormark{4}Laboratory of Photonics, Tampere University, FI-33014 Tampere, Finland\\
\authormark{5}The Institute of Optics and Department of Physics and Astronomy, University of Rochester, Rochester, New York 14627, USA\\

}

\email{\authormark{*}msaad009@uottawa.ca} %% email address is required

% \homepage{http:...} %% author's URL, if desired

%%%%%%%%%%%%%%%%%%% abstract %%%%%%%%%%%%%%%%
%% [use \begin{abstract*}...\end{abstract*} if exempt from copyright]

\begin{abstract}
Multiresonant metasurfaces could enable many applications in filtering, sensing and nonlinear optics. However, developing a metasurface with more than one high-quality-factor or high-$Q$ resonance at designated resonant wavelengths is challenging. Here, we experimentally demonstrate a plasmonic metasurface exhibiting different, narrow surface lattice resonances by exploiting the polarization degree of freedom where different lattice modes propagate along different dimensions of the lattice. The surface consists of aluminum nanostructures in a rectangular periodic lattice. The resulting surface lattice resonances were measured around 630 nm and 1160 nm with $Q$-factors of $\sim$50 and $\sim$800, respectively. The latter is a record-high plasmonic $Q$-factor within the near-infrared type-II window. Such metasurfaces could benefit applications such as frequency conversion and all-optical switching.
\end{abstract}

%%%%%%%%%%%%%%%%%%%%%%%%%%  body  %%%%%%%%%%%%%%%%%%%%%%%%%%
Resonant metasurfaces promise to enable free-space photonic applications in nanoscale thin flat optical devices. Thanks to their strong resonance enhancement characteristics, plasmonic lattice metasurfaces formed by metal nanostructures are considered to be a strong candidate for applications such as sensing, spectroscopy, and lasing ~\cite{Spackova2016,wang2019manipulating,Kravets2018}. Among those applications, some specific processes may involve two or more frequencies, particularly nonlinear optical processes such as harmonic generation, frequency up- and down-conversion, cross-phase modulation or ultrafast all-optical switching~\cite{boyd2020nonlinear}. Strongly resonant responses like plasmonic resonances could boost the efficiency of nonlinear optical processes without  requiring any phase-matching between the input and output waves~\cite{Kauranen2012,bin2020hyperpolarizability}. Thus, an implementation of multiresonant plasmonic metasurfaces could dramatically enhance the efficiency of applications involving nonlinear optical phenomena~\cite{Butet2015}.

%\caption{(a--b) A 2D illustration and a focused-ion beam micrograph of the metasurface. (c) Simulated and (d) experimentally measured normalized transmission spectra of the polarization-dependent multiresonant LSPRs and SLRs in a plasmonic metasurface consisting of V-shape Aluminum nanostructures array. The inset plots of (c) and (d) demonstrate the simulated and experimentally measured normalized transmission spectra of the near-infrared ($\sim$ 1155 nm to 1160 nm wavelength) SLRs, which are significantly narrowband with very high-$Q$ values ($\sim$ 700 to 800).}

%Surface plasmons are excitations generated at the interface of a metal-dielectric surface when under illumination. Nanoscale metallic structures can confine the energy of surface plasmons in localized surface plasmon resonances (LSPRs)~\cite{Maier2007}. 

Under optical illumination, metal nanostructures naturally exhibit strongly localized surface plasmon resonances (LSPRs)~\cite{Meinzer2014}. However, a longstanding issue with metal nanostructures are their high absorptive and radiative losses, which result in the swift decay of excitations associated with the LSPRs. Whereas absorptive losses are inherent to metals, it is possible to suppress the radiative or scattering losses by engineering periodic arrays of metal nanostructures to support plasmonic surface lattice resonances (SLRs) with longer lifetimes~\cite{Kravets2018,cherqui2019plasmonic}. Such SLRs originate from the collective coupling of every particle in a lattice and suppress the scattered losses associated with the individual particles. Thus, the resulting resonances can have significantly high quality-factors ($Q>2000$~\cite{bin2021ultra}), appearing at the wavelength near the diffraction edge of the lattice periodicity. Under normal illumination, the spectral position of the SLRs can be defined by the product of the periodicity, $P$ of the particles towards its radiating direction and the refractive index $n$ of the particles' surrounding background medium ($\lambda_\mathrm{SLR} \approx nP$)~\cite{bin2021ultra,Khlopin2017}.

% Now that we have SLRs, what can we do?
% Paragpraph to explain the issue we specifically are solving, about SLRs
%Recent studies suggest that SLRs can simultaneously provide the benefits of both plasmonic metasurfaces (strong field confinement and large nonlinearities) as well as dielectrics (high-$Q$ and long interaction time). These properties have proven to improve performance for applications in nonlinear optics~\cite{Klein2006, Czaplicki2018, Michaeli2017a, Kauranen2009}, nanolasers, biosensing performance. \OR{Add citations specific to SLR for these applications.} Nevertheless, 

 %\begin{figure}[ht!]
\begin{figure}[htbp]
\centering
\includegraphics[width=100mm]{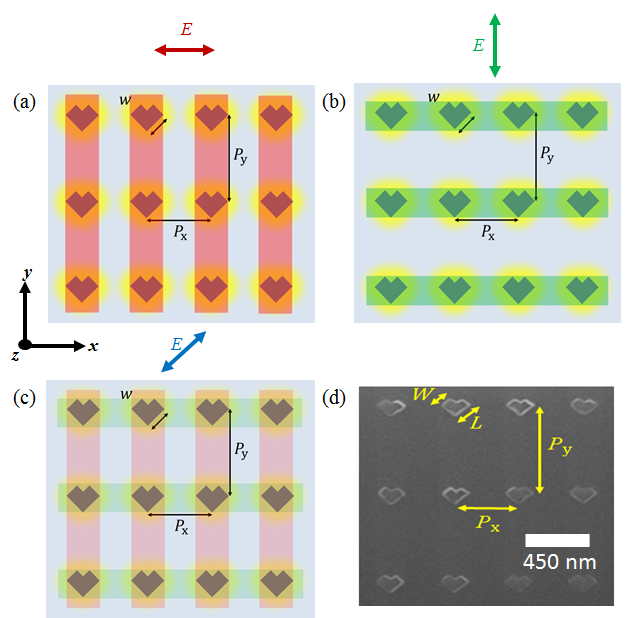}
\caption{2D illustrations of plasmonic metasurface. The normally incident light is (a) polarized along the $x$-axis to excite SLRs alongside the periodicity $P_{y}$ (red color), (b) polarized along the $y$-axis to excite SLRs alongside the periodicity $P_{x}$ (green color) and (c) diagonally polarized to simultaneously excite SLRs alongside both directions (red and green colors). The yellow color at the edge of the individual V-shape nanostructures represents The LSPRs. (d) A focused-ion beam micro-graph of a fabricated metasurface.}
\label{fig:Schematic}
\end{figure}

As such, SLR modes are generated in the form of in-plane waves, oscillating orthogonally to the polarization direction of incident light. Hence, in a 2D plasmonic metasurface array with a rectangular lattice formation, it is possible to generate two different SLR modes along two separate orthogonal directions~\cite{Huttunen2019}. Such lattice modes owing to orthogonally polarized incident waves may be useful for many nonlinear optical processes, such as cross-phase modulation or two-beam coupling, which depend on more than one input field. %Nonetheless, other nonlinear optical processes like second- and third-harmonic generation require only a single input beam to generate one output harmonic signal having the same polarization state. In the later case, a technique with which one can simultaneously excite more than one lattice plasmon mode is desired.

%Thus, the situation demands a technique to simultaneously excite more than one lattice plasmon modes just by only one incident polarization state. 
%In this Letter, we report the superposition of dual-localized surface plasmon and dual-lattice plasmon modes in different wavelengths which are co-excited by the diagonally polarized incident light. Here, both orthogonal components, $i.e.$ $x$- and $y$- components, of the diagonally polarized light contribute in the generation of the different types of plasmon resonances simultaneously. 

 %\begin{figure}[ht!]

In this Letter, we report the observation of multiple LSPRs and SLRs at different wavelengths in the same metasurface. The resonant wavelengths of the SLRs can be selected through the careful choice of lattice geometry. The different resonances can be isolated by selecting a given linear polarization of the probing light or they can be excited simultaneously by employing diagonally polarized light.

To properly design a metasurface with multiple resonances in both the visible and infrared regime, here we consider the behaviour of the constituent materials in the spectral range from 400 nm to 1300 nm. In noble plasmonic metals like gold and silver, the interband transition occurs in the visible wavelength regime~\cite{Maier2007}. This transition causes such metals to absorb most of the visible light, and to lose their capability to support plasmon oscillations in the shorter ultraviolet wavelength regime. However, unlike noble metals, the interband transition in aluminum appears in the near-infrared regime (around $\lambda=850 \mathrm{~nm})$. Furthermore, because of the electronic band structure of the Al, the interband transition is quite narrow~\cite{Gerard2015}.
Hence, aluminum retains its metallicity at shorter wavelengths compared to gold or silver. Subsequently, aluminum nanostructures can exhibit LSPRs and thus can efficiently scatter light in the visible or ultraviolet spectral ranges~\cite{Halas2014}. Inspired by this fact, aluminum nanostructures have been recently adopted to demonstrate SLRs in periodic metasurfaces with applications in SHG and nanolasing~\cite{Khlopin2017,Huttunen2019,wang2019manipulating}. It was also revealed that aluminum possesses comparatively larger nonlinear optical coefficient than gold or silver~\cite{Metzger2015,Hulst2011}. We therefore elect to have our metasurface composed of aluminum nanoparticles cladded in a transparent, fused silica substrate.

To create polarization-dependent LSPRs, we choose to use V-shape nanostructures, which can individually exhibit two LSPRs at different wavelengths~\cite{Kauranen2012}. We design and fabricate periodic right--angle V-shape aluminum metal nanostructures. The fabricated dimensions are: length $L$ $\approx$ 110--130 nm, width $W$ $\approx$ 70--80 nm, and thickness $t$ $\approx$ 30 nm. The 2D illustration of the designed metasurface is demonstrated in Fig.~\ref{fig:Schematic}(a--c). Here we depict the LSPR in yellow, the SLR mode for $x$-polarization in red, and the SLR mode for $y$-polarization in green in Fig.~\ref{fig:Schematic}(a--b), respectively. In Fig.~\ref{fig:Schematic}(c), we show that both SLRs can be excited simultaneously using diagonally polarized illumination. A focused-ion beam micro-graph of the fabricated array is depicted in Fig.~\ref{fig:Schematic}(d). The periodic structured particles form a rectangular lattice in the $xy$ plane (periodicity $P_\mathrm{x}$ $\approx$ 445 nm and $P_\mathrm{y}$ $\approx$ 790 nm) inside a fused silica substrate (refractive index $n$ = 1.46). 

The sample fabrication and the experiment technique are similar to those of Ref.~[\citenum{bin2021ultra}]. We fabricate metasurfaces using a standard metal lift-off process. On top of a fused silica substrate, we deposit a silica undercladding layer using sputtering. Next, we define the pattern of the nanostructure arrays using electron-beam lithography in a positive tone resist bi-layer with the help of a commercial conductive polymer. To correct for the nanostructure corner rounding, we design the mask using shape-correction proximity error correction. After the development, we deposit an aluminum layer using thermal evaporation followed by the lift-off process. We deposit a final protective silica cladding layer using sputtering. To make sure that the environment surrounding the metasurface is completely homogeneous, we sputter the initial and final silica layers using the same tool under the same conditions. Before the characterization, we cover the surface of the device in index-matching oil to make the metasurface surrounding medium homogeneous. %To minimize substrate-related etalon fringes, we also coated the backside of the silica substrate with an anti-reflective coating.

In the experiment, we use a normally incident collimated light beam from a broadband supercontinuum laser source (spectral range $\lambda$ = 470 to 2400 nm) to flood-illuminate all of the metasurface arrays in the sample. We control the incident polarization using a broadband linear polarizing filter. We observe the image of the light transmitted by the metasurface by a lens with a focal length $f$ = 35 mm and by placing a 100 $\mu$m pinhole in the image plane to collect the image of the desired array. We use a large-core (400 $\mu$m) multimode fiber to collect the transmitted light from the sample metasurface array and pass it to an optical spectrum analyzer (OSA). The resolution of the OSA is set to 0.01 nm. The OSA is used to analyze the measured light which is normalized to a background trace of the substrate without aluminum nanostructures.

\begin{figure*}[htbp]
\centering
\includegraphics[width=130mm]{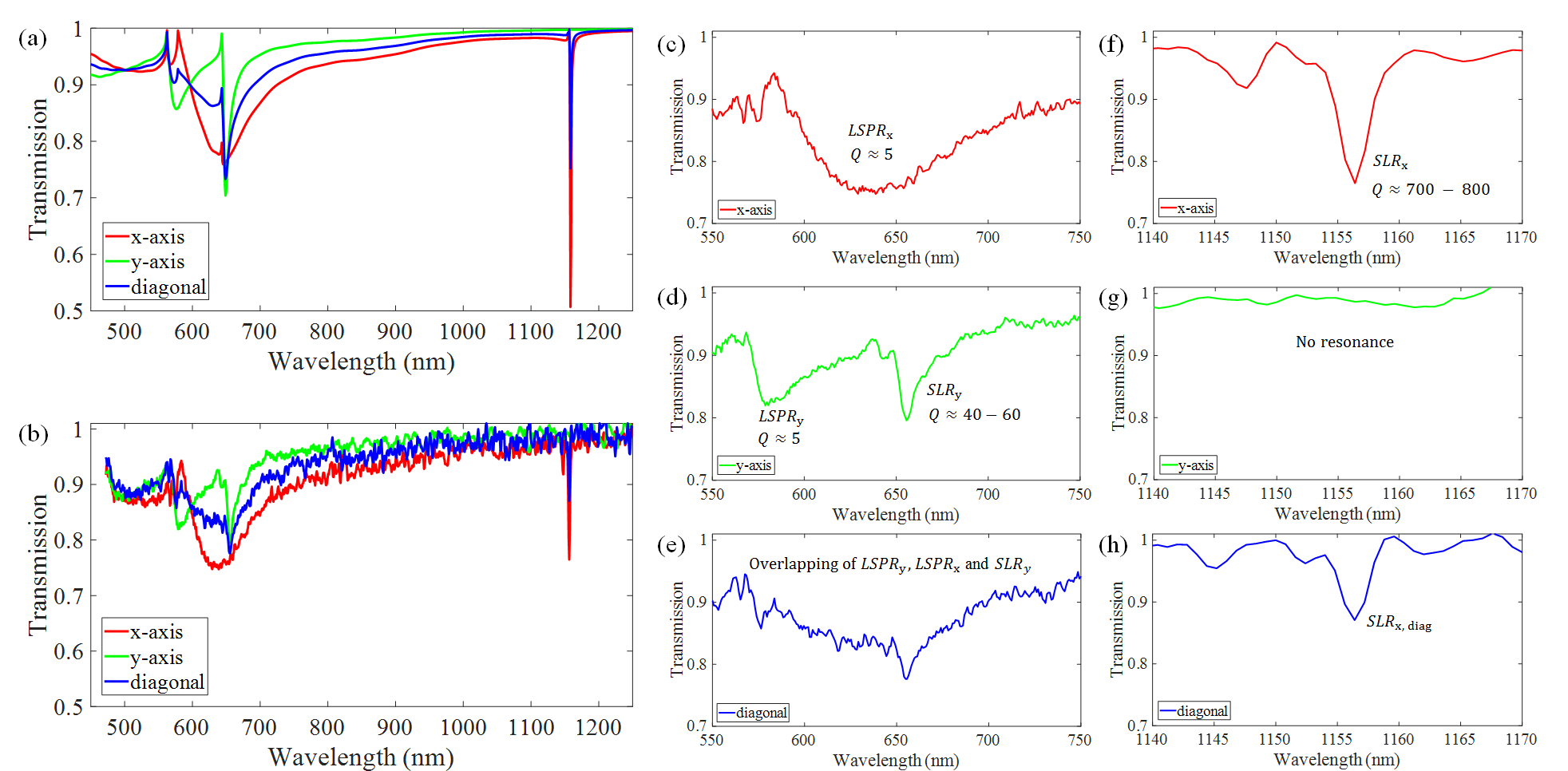}
\caption{(a) Simulated and (b) experimentally measured normalized transmission spectra of the polarization-dependent ($x$-axis, $y$-axis and diagonal) multiresonant LSPRs and SLRs in the plasmonic metasurface shown in Fig.~\ref{fig:Schematic}(a--d). Diagonally polarized light simultaneously excites the LSPRs and SLRs of both dimensions, enabling simultaneous SLRs around 650 nm and 1150 nm. The zoomed-in measured spectra are shown in (c--e) for the visible resonances and in (f--h) for the near-infrared resonances for $y$-axis and diagonal polarization, respectively.}
\label{fig:Spectra}
\end{figure*}

 Fig.~\ref{fig:Spectra}(a) shows the simulated transmission spectra of our designed metasurface for different polarization states. The simulation was performed using Lumerical FDTD. We also measure the normalized transmission spectra of our fabricated metasurface illuminating it by a normally incident collimated beam generated from a broadband supercontinuum source. Fig.~\ref{fig:Spectra}(b) shows the measured normalized transmission spectra for different polarization states which are in good agreement with the simulated results, shown in Fig.~\ref{fig:Spectra}(a). For convenience, the resonances in the visible and infrared are presented separately in Fig.~\ref{fig:Spectra}(c--h) for different polarization states. It is evident in the zoom-in spectra, that in the visible regime, the x-polarized LSPR (Fig.~\ref{fig:Spectra}(c)) overlaps with the y-polarized LSPR and SLR (Fig.~\ref{fig:Spectra}(d)) under diagonally polarized excitation (Fig.~\ref{fig:Spectra}(e)). We note that the resonance strength is halved from its original value as conceptually predicted in Fig.~\ref{fig:Schematic}(c) and simulated in  Fig.~\ref{fig:Spectra}(a). Such overlapping causes the resonances in the visible regime to superimpose with each other and thus form a modified spectral lineshape, as depicted in Fig.~\ref{fig:Spectra}(e). The $Q$-factor of the LSPR for the $x$-polarization is around 5, whereas for the $y$-polarization, the LSPR $Q$-factor is around 25 and the SLR $Q$-factor is around 50.
 
As expected, the SLR for the $x$-polarization in the infrared regime (shown in Fig.~\ref{fig:Spectra}(f)) emerged at far spectral distance red-shifted from all other resonance features in the visible regime. From different samples of our designed metasurface, we find the $Q$ value for this infrared SLR to vary between 700 and 820. Although high-$Q$ SLR feature completely vanishes under the $y$-polarized excitation (see Fig.~\ref{fig:Spectra}(g)), it can emerge without any spectral modification for any other polarization states with reduced strength. Thus, as expected, we observe the re-appearance of the high-$Q$ SLR in Fig.~\ref{fig:Spectra}(h) for the diagonal polarization.  %one pure SLR mode in the near infrared regime only for the $x$-axis and diagonal polarization. The strong LSPR mode around 630 nm and the weak SLR mode around 650 nm wavelengths for the $x$-axis polarization overlap so intimately (see Fig.~\ref{fig:Spectra}(c)) that is not clearly observable in the measured curve (Fig.~\ref{fig:Spectra}(d)). Moreover, that weak SLR mode around 650 nm wavelength corresponds to the periodicity $P_\rm{x}$ which becomes strong for the $y$-axis and diagonal polarization. In the opposite way, the strong LSPR mode around 630 nm gradually becomes weaker. This LSPR mode around 630 nm originally corresponds to the individual V-shape nanostructures for $x$-axis polarization. 

\begin{figure*}[htbp]
\centering
\includegraphics[width=130mm]{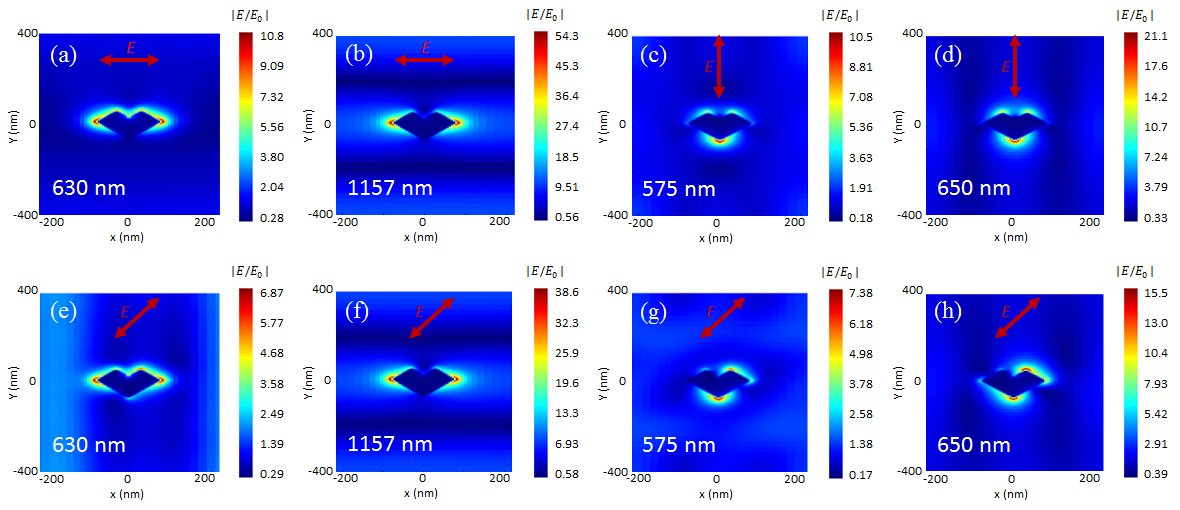}
\caption{Normalized electric field distributions of the SLR modes for the $x$-polarization at (a) 630 nm and (b) 1157 nm, $y$-polarization at (c) 575 nm and (d) 650 nm, and diagonal polarization at (e) 630 nm, (f) 1157 nm, (g) 575 nm and (h) 650 nm.}
\label{fig:E_map}
\end{figure*}

Next, we investigated the impact of the polarization state rotation on the in-plane electric-field distribution inside the metasurface for all the resonances we discussed above (Fig.~\ref{fig:E_map}). The simulated normalized electric-field distribution corresponding to the LSPR around $\lambda =$ 629 nm is presented in Fig.~\ref{fig:E_map}(a) and the SLR around 1157 nm is presented in Fig.~\ref{fig:E_map}(b) for the $x$-polarization. These figures depict that the horizontally excited LSPR mode is only localized near the individual nanostructures; however, the delocalized SLR mode forms a delocalized diffraction grating-like standing wave, extended along the $y$-axis orthogonal to the polarization direction. The electric field strength of the SLR is significantly higher than that of its LSPR counterpart, as indicated by the color bars in Fig.~\ref{fig:E_map}(a) and Fig.~\ref{fig:E_map}(b), respectively. Such a large field enhancement in the SLR is the outcome of the scattering loss reduction by trapping the energy of scattered light near the diffraction order. The field distributions for these modes under $y$-polarized illumination are presented in Figs.~\ref{fig:E_map}(c-d). These modes are similar to their $x$-polarized counterparts, except they are pointed along the $y$-direction, and feature smaller field enhancements due to a lower value of the associated $Q$-factor. %are similar to their counterparts for the $x$-polarization but with three exceptions: (1) the LSPR mode is excited vertically, (2) the SLR mode now forms the diffraction grating-like standing wave towards the $x$-axis orthogonal to the $y$-polarization state, and (2) the SLR field enhancement comparing to the LSPR is not very large due to its relatively low-$Q$ nature demonstrated in Fig.~\ref{fig:Spectra}(d), although its $Q$-factor is still large enough than the $Q$ values of the typical LSPRs demonstrated in Fig.~\ref{fig:Spectra}(c) and Fig.~\ref{fig:Spectra}(d), respectively. 

We now turn our attention to the electric-field distribution under diagonal polarization. The field distribution in Fig.~\ref{fig:E_map}(e) depicts the LSPR field under diagonally polarized light with a wavelength centered at 630 nm. Here an SLR-like grating mode can be observed to form along the edges of the unit cell. The generation of this SLR-like feature around the LSPR center wavelength occurs because of the influence of $y$-polarized SLR mode around 650 nm (Fig.~\ref{fig:E_map}(d)). This phenomenon represents the hybridization between the $x$-polarized LSPR and the $y$-polarized SLR under diagonally polarized light. Comparing the scales between Fig.~\ref{fig:E_map}(a) and Fig.~\ref{fig:E_map}(e), we see that the overall field strength of the original LSPR for the $x$-polarization is slightly reduced in the case of hybridized mode around 630 nm by the off-axis diagonal polarization state.

%alongside a newly emerged delocalized SLR-like diffraction grating towards the $x$-axis for the $y$-component of the diagonal polarization at 648.8 nm. This feature in fact confirms the spectral modification and hybridization of the LSPR$_\rm{x_{diagonal}}$ for the $x$-polarization at 650 nm by the SLR$_\rm{y_{diagonal}}$ for the $y$-polarization at 648.8 nm shown earlier in Fig.~\ref{fig:Spectra}(e). Nevertheless, such resonance mode overlapping causes a noticeable field enhancement, mostly surrounding the individual nanostructures' edge, indicated by the difference of the values of the color bars' reading from both Fig.~\ref{fig:E_map}(a) and Fig.~\ref{fig:E_map}(e).

In contrast, Fig.~\ref{fig:E_map}(f) shows no deviation of the standalone high-$Q$ SLR field distribution around 1157 nm for the diagonal polarization from its $x$-polarization counterpart in Fig.~\ref{fig:E_map}(b), except the reduction of the field strength values indicated by the color bars. This feature confirms the behavior previously observed in Fig.~\ref{fig:Spectra}(h), which depicts no spectral modification of the sharp infrared SLR in Fig.~\ref{fig:Spectra}(f) due to the lack of any other resonance nearby in Fig.~\ref{fig:Spectra}(g). Next, we see that the LSPR field distribution for the $y$-polarization around 575 nm in Fig.~\ref{fig:E_map}(c) is also modified by the diagonally polarized light in Fig.~\ref{fig:E_map}(g). Such a modification occurs because of the hybridization of this LSPR for the $y$-polarization around 575 nm with the LSPR for the $x$-polarization around 630 nm. Contrary to the LSPR around 630 nm, this short wavelength modified LSPR around 575 nm is positioned relatively far from the SLR around 650 nm. Thus, the corresponding field distribution in Fig.~\ref{fig:E_map}(g) does not show a pronounced SLR-like grating. 

Lastly, the field distribution in Fig.~\ref{fig:E_map}(h) depicts an SLR-like field distribution, just like in Fig.~\ref{fig:E_map}(d). These observations demonstrate that the field distribution of the localized plasmon-like modes are dominated by the polarization direction; conversely, the SLR modal distribution is dictated by the excitation wavelength.

%Lastly, the field distribution in Fig.~\ref{fig:E_map}(h) depicts that, the delocalized grating-like SLR field distribution for the $y$-component of the diagonal polarization around 650 nm does not deviate from the field distribution for the purely $y$-polarized light in Fig.~\ref{fig:E_map}(d). However, the hybridized localized field surrounding the individual nanostructures' edge now tends to be stretched towards the diagonal direction. As expected, the relevant color bars again tells that, the diagonal polarization reduces the field strength here as like the other cases, too.

In summary, we have experimentally demonstrated a technique to simultaneously excite cross-polarized dual LSPR and SLR modes in a nanostructured periodic metasurface via diagonally polarized normally incident light. Here, two different orthogonal periodicities $P_{x}$ and $P_{y}$ correspond to two separate SLRs centered at wavelengths $\lambda$ = 650 nm and $\lambda$ = 1157 nm, respectively. These two SLRs can be tuned independently by modifying $P_{x}$ and $P_{y}$. We showed that these two SLRs can be excited simultaneously using light that is diagonally polarized with respect to the rectangular lattice of the metasurface; it should also be doable using circular polarized light. We analyzed their resonance characteristics and the relevant electric field distribution of the generated modes. In addition, we observed an unprecedentedly large $Q$-factor for the SLR in the infrared type-II regime ($Q$ $\approx$ 800 around $\lambda$ = 1157~nm). We believe that the $Q$ of the SLRs in the visible regime could be further enhanced (up to 200~\cite{Huttunen2019}) by further optimizing the dimensions of the individual plasmonic nanostructures, the periodicity alongside the $x$-axis, and by enlarging the metasurface area~\cite{bin2021ultra,Huttunen2019}. We hope, our multiband high-$Q$ plasmonic metasurface will pave the way for efficient nonlinear optical processes in flat photonic devices.

%\section{Backmatter}

%Backmatter sections should be listed in the order Funding/Acknowledgment/Disclosures/Data Availability Statement/Supplemental Document section. An example of backmatter with each of these sections included is shown below.

\begin{backmatter}
%\bmsection{Funding}
%Content in the funding section will be generated entirely from details submitted to Prism. Authors may add placeholder text in the manuscript to assess length, but any text added to this section in the manuscript will be replaced during production and will display official funder names along with any grant numbers provided. If additional details about a funder are required, they may be added to the Acknowledgments, even if this duplicates information in the funding section. See the example below in Acknowledgements.

\bmsection{Acknowledgments}
The authors thank Iridian Spectral Technologies Ltd, Ottawa, ON, Canada for help in cladding the silica superstrate on top of the metasurface substrate for this work.

\bmsection{Disclosures}
The authors declare no conflicts of interest.

\medskip

%\noindent ABC: 123 Corporation (I,E,P), DEF: 456 Corporation (R,S). GHI: 789 Corporation (C).

\medskip

\end{backmatter}

\bibliography{L_Shape_SLRs}

%%%%%%%%%% If preparing manually:
% \begin{thebibliography}{1}
% \newcommand{\enquote}[1]{``#1''}

% \bibitem{Zhang:14}
% Y.~Zhang, S.~Qiao, L.~Sun, Q.~W. Shi, W.~Huang, L.~Li, and Z.~Yang,
%   \enquote{Photoinduced active terahertz metamaterials with nanostructured
%   vanadium dioxide film deposited by sol-gel method,}
%   {\protect\JournalTitle{Optics Express}} \textbf{22}, 11070--11078 (2014).

% \bibitem{OSA}
% {Optical Society}, \enquote{{OSA Publishing},}
%   \url{http://www.osapublishing.org}.

% \bibitem{FORSTER2007}
% P.~Forster, V.~Ramaswamy, P.~Artaxo, T.~Bernsten, R.~Betts, D.~Fahey,
%   J.~Haywood, J.~Lean, D.~Lowe, G.~Myhre, J.~Nganga, R.~Prinn, G.~Raga,
%   M.~Schulz, and R.~V. Dorland, \enquote{Changes in atmospheric consituents and
%   in radiative forcing,} in \enquote{Climate Change 2007: The Physical Science
%   Basis. Contribution of Working Group 1 to the Fourth assesment report of
%   Intergovernmental Panel on Climate Change,}  S.~Solomon, D.~Qin, M.~Manning,
%   Z.~Chen, M.~Marquis, K.~B. Averyt, M.~Tignor, and H.~L. Miler, eds.
%   (Cambridge University Press, 2007).

% \end{thebibliography}

\end{document}